# High-energy interactions in Kinetic Inductance Detectors arrays

A. D'Addabbo*[a,b], M. Calvo[a], J. Goupy[a], A. Benoit[b], O. Bourrion[c], A. Catalano[c], J. F. Macias-Perez[c] and A. Monfardini[a]

[a]Institut Néel and Université Joseph Fourier, CNRS, BP 166, 38042 Grenoble, France; [b]Università Sapienza, Piazzale Aldo Moro 5, 00185, Rome, Italy ; [c]Laboratoire de Physique Subatomique et de Cosmologie (LPSC), CNRS and Université de Grenoble, France.

## ABSTRACT

The impacts of Cosmic Rays on the detectors are a key problem for space-based missions. We are studying the effects of such interactions on arrays of Kinetic Inductance Detectors (KID), in order to adapt this technology for use on board of satellites. Before proposing a new technology such as the Kinetic Inductance Detectors for a space-based mission, the problem of the Cosmic Rays that hit the detectors during in-flight operation has to be studied in detail. We present here several tests carried out with KID exposed to radioactive sources, which we use to reproduce the physical interactions induced by primary Cosmic Rays, and we report the results obtained adopting different solutions in terms of substrate materials and array geometries. We conclude by outlining the main guidelines to follow for fabricating KID for space-based applications.

**Keywords:** KID, Cosmic Rays, phonon propagation, space missions.

## 1. INTRODUCTION

Over the last decade KID have been undergoing a rapid development, and their performances at millimeter wavelength are now comparable to other state-of-the-art bolometers. Thanks to their intrinsic multiplexability in frequency domain, KID are naturally suited for hundred-pixels cameras operating at very low temperatures. The first on-sky arrays counting hundreds of detectors have been successfully deployed by the NIKA collaboration, which has demonstrated that KID are ready for ground based applications[1]. Nevertheless, suitability of this technology for satellite based telescopes is still to be proved. One of the most challenging problems for space borne experiments is that the detectors are exposed to an intense flux of high energy particles, referred to as Cosmic Rays[2] (CR). The glitches caused by CR hits can saturate the detectors making the corresponding data useless and leading to the loss of an important fraction of the observing time. In addition, small amplitude glitches could remain hidden in the noise introducing an additional non-gaussian component. A proper understanding of the glitches physical origin is therefore a key point for the next generation of precision cosmology space missions.This document describes the current knowledge we have of such events, the system that has been developed to study them more in depth, and the first results obtained.

## 2. THE COSMIC RAYS ISSUE

Any instrument working outside the Earth's protecting atmosphere is constantly exposed to an intense flux of CR. These high-energy particles are produced by the Sun and by other galactic sources, and are composed primarily by protons (≈90%), with an important contribution of helium nuclei (≈10%), and a few heavier nuclei (≈1%) and electrons (< 1%). The actual intensity of this flux, and the consequent rate of hits on the instrument, is modulated by the solar activity. The particles having sufficient energy to penetrate the shields can reach the detectors sensitive part and give an unwanted signal. The energy deposited by an event is given by the Bethe-Bloch formula

$$-\frac{dE}{dx} = \frac{4\pi}{m_e c^2} \frac{nz^2}{\beta^2} \left(\frac{e^2}{4\pi\varepsilon_0}\right)^2 \left[ ln\left(\frac{2m_e c^2 \beta^2}{I(1-\beta^2)}\right) - \beta^2 \right] \qquad (1)$$

*antonio.d-addabbo@grenoble.cnrs.fr

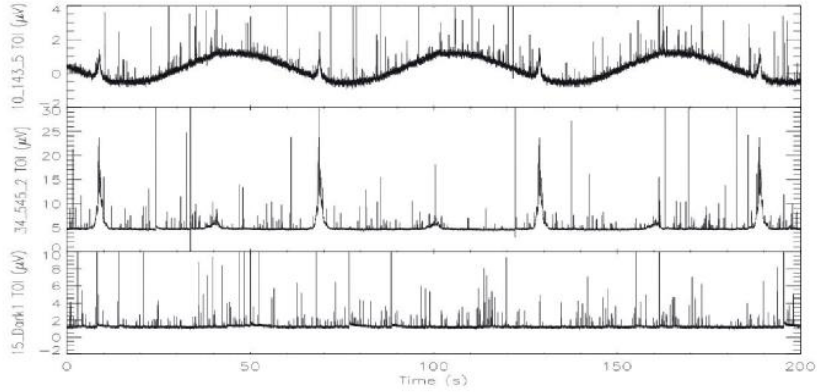

Figure 1. Example of Planck time traces. Raw TOIs for three bolometers, 143GHz (top), 545GHz (middle), and a Dark1 bolometer (bottom) over three rotations of the spacecraft (at 1rpm). The typical maximum amplitude of a spike is between 100 and 500 mV depending on the bolometer. The typical rate is one event per second per detector.

where $n$ is the electron number density of the target material, $I$ is its mean excitation potential, and $z$ and $\beta c$ are the charge and the speed of the particle. Considering that the peak of the CR spectrum is at about 200 MeV, this corresponds roughly to a deposited energy of 0.5 keV per μm traversed for a silicon target.

The energy released by a CR hit is followed by a lapse of time during which the hit detector is saturated and therefore "blind" to the incoming scientific signal. The length of this dead-time depends on the intrinsic time constant of the sensor, which in turn depends on the physical processes that convert the deposited energy in the variation of a measurable physical quantity.

## 2.1 Bolometer response to Cosmic Rays

Bolometers are thermal detectors, in which the optical and thermal power is first converted into thermal phonons and then transduced by a suitable thermistor into a measurable electrical signal. As far as the interactions with CR are concerned, this has two main consequences: first, all the energy released by the CR can be sensed, and second, bolometers exhibit relatively slow time constants, typically of 5 ms or more. This response time is primarily determined by the thermalization of the energy released by the incident radiation, which essentially depends on the thermal contact between the thermistor and the temperature bath. In order to match high sensitivity requirements for space born missions, this contact has to be very weak, resulting in slow response times. These span from tens of milliseconds to several seconds depending on the type of interaction, on the hit point and on the propagation mechanism of the deposited energy to the thermistor.

## 2.2 The Planck mission experience

The Cosmic Rays impacts have been one of the main problems during the reduction of the data taken by the Planck satellite mission[3]. The total amount of data lost due to CR impacts calculated by the Planck collaboration varies between 8% and 20%, larger than what was expected as a consequence of the coupling between the absorber and the CRs at the lagrangian point L2 (see Figure 1). Due to the instrument shielding, only protons with energy exceeding 39 MeV are able to reach the focal plane and induce a signal in the detectors. Considering the CR spectrum[4], this cut-off energy would translate in a rate of about 5events/min/mm$^2$, modulated by the solar activity. Planck detectors were spider-web bolometers, in which the sensitive element, that is the NTD thermistor, had a very small surface area (0.03mm$^2$ for normal incidence). If one assumes that only the impacts on the thermistor itself or on the spider-web structure (around 0.1mm$^2$) are able to cause a glitch in the data, this means that the expected event rate on Planck's detectors was roughly 0.5ev/min.

The rate actually observed in flight was 100 times higher, of the order of 1 Hz[5]. This excess of glitches can be accounted for if also the CR interacting with the silicon die can be sensed. In fact, each CR hit creates a cascade of high-energy phonons: a part of such phonons, referred to as *ballistic phonons,* can propagate rapidly (≈ mm/μs) and over large distances in the substrate, being reflected at each interface[6]. Ballistic phonons then decay within hundreds of microseconds into thermal ones. Several laboratory and accelerator experiments following the Planck launch confirmed this[7], demonstrating that ballistic phonons can transfer part of the energy released by a CR hitting the silicon die through the spider-web and into the bolometer absorber and NTD thermometer. Here the energy can be thermalized and induce a signal.

The spider-web structure is therefore not insulating enough to suppress the unwanted signals coming from the silicon die: this means that, when estimating the CR events rate, also the surface of silicon side-structure must be taken into account, which explains the increased observed rate and therefore the higher percentage of lost data. In Planck, the silicon die has an area between 0.4 and 0.8 cm$^2$, depending on the working frequency. If, as it has been suggested, every cosmic particle hitting the die is seen by the thermistor (through ballistic phonons mediation), from the total rate of glitches observed in Planck ($\approx$ 1 Hz) we could estimate the integrated rate of CR per unit surface of 1event/s/5·10$^{-5}$ m$^2$ = 2 ·10$^4$ events/s/m$^2$.

## 3. ESTIMATING COSMIC RAYS EFFECT ON KID

The operating principle of Kinetic Inductance Detectors is completely different from that of bolometers. The KID are *pair-breaking detectors*, which are sensitive to variations in the Cooper pair density of the superconductor. A signal can be induced into the detector only by particles able to release an energy greater than the binding energy of each Cooper pair

$$2\Delta = 3.5\ k_B T_c \qquad (2)$$

where $k_B$ is the Boltzmann constant, and $\Delta$ and $T_c$ are the energy gap and the critical temperature of the superconductor, respectively. This makes the KID insensitive to phonons of energy below $2\Delta$. Therefore only high energy phonons can cause a signal, and as soon as these decay to energy levels below the gap, they become effectively invisible from the KID point of view. For thermal phonons the energy down-conversion occurs over very fast time scales, and they can only cover short distances before becoming undetectable. On the other hand, ballistic phonons can propagate over long distances before losing their energy through scatterings with the die edges or with the impurities present into the substrate lattice[8]. This can lead to a glitch even in detectors far from the impact point. The energy down-conversion of the phonons depends on the geometry and on the properties of the substrate. Since the decay of the thermal phonons below the gap is very fast, the *phonon lifetime* that we observe is essentially due to ballistic ones. This is described by the time scale $\tau_{ph}$, typically of order of hundreds of microseconds for silicon dies[9].

The time evolution of a CR event in a KID is also determined by the quasi-particle recombination time, or *quasi-particle lifetime*, $\tau_{qp}$. This describes the average elapsed time between a quasi-particle creation and its recombination into a Cooper pair. Since two quasi-particles have to interact to form a pair, the recombination is faster when their density is higher, so that $\tau_{qp}$ increases when lowering the temperature of the device[10].

The time constant governing the KID behaviour can be easily measured, although discriminating between the effect of quasi-particles and phonons lifetimes is not straightforward[11]. It has been shown that for very low-temperature aluminum KID fabricated on silicon substrate, this time constant is of the order of fractions of millisecond[9], much faster than in the case of bolometers, which fundamental thermal time constant is of a few milliseconds . This means that for the same CR event rate, less data are lost when using KID than when using bolometers. To make a preliminary estimate, we can compare the typical expected performance of a KID with respect to a Planck bolometer. The surface area of a typical NIKA-like array ( pixels) of LEKID operating at millimeter wavelength[12] is roughly 13 cm$^2$ = 1.3·10$^{-3}$ m$^2$. So, if the Planck hypotheses and the rate of cosmic particle per unit area they imply are correct, the total rate of particles hitting the array should be, in the case of KID, of the order of 20 times higher than Planck, i.e. 20 events/sec. Assuming KID with a time constant of 0.1 ms, this translates in a data loss analogue to that of the Planck mission, which resulted in a reduction of the observing time of about 15%. We can therefore already estimate that a hundred-pixel array of KID fabricated on a single bulk silicon wafer should not suffer from data loss higher than a single Planck pixel.

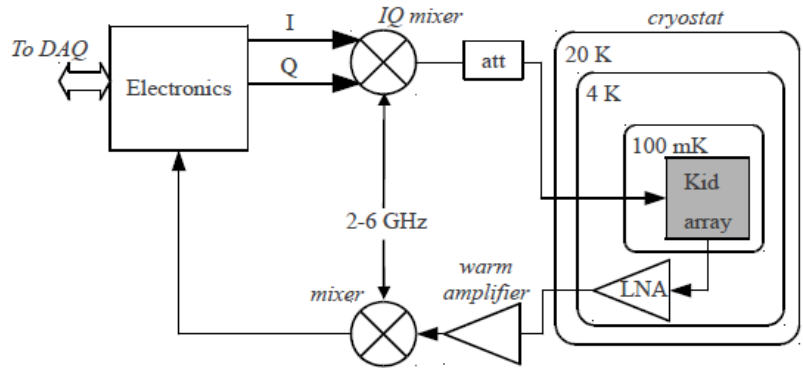 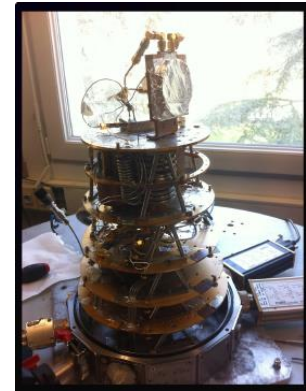

Figure 2. On the left, a schematic representation of the read-out setup. The electronics DACs generate the in-phase (I) and quadrature-phase (Q) signals, they are then up-converted to GHz frequency with an IQ mixer device, attenuated before reach the detectors array. The output signal passes through a Low Noise Amplifier at 4K, eventually further amplified at room temperature before the down-conversion to electronics bandwidth. Finally the signal is red by ADC and send to Data Acquisition (DAQ) system. On the right, an image of the dilution cryostat of our test-bench. Each plate is cooled down to temperature decreasing upwards, reaching the base temperature at the top, where two arrays are mounted and ready to be tested.

## 4. DEDICATED TEST-BENCH FOR HIGH ENERGY PARTICLES

In order to better understand and model the evolution and propagation of the phonons in the substrate, it is worth to carry out a detailed study of the effect of high energy impacts on Kinetic Inductance Detectors. We have developed a dedicated test-bench to test the technologies currently in use and the different solutions that can be adopted to mitigate the effect of CR impacts. In the following we will give a brief description of the main components of our experimental setup.

### 4.1 Cryostat

The cryostat is a two-stage system with a base temperature of the system is 100mK. Liquid helium is used to reach 4K, while the cooler stage is provided by a $^3$He/$^4$He dilution refrigerator. The shields at the various temperatures are equipped with optical windows in order to be able to perform measurements under varying optical loads. The cryostat has two independent RF channels, each one equipped with a cold low-noise amplifier optimized for the working frequencies of our detectors.

### 4.2 Fast readout electronics

The readout electronics is based on a FPGA board similar to the one used for the NIKA experiment, but modified in order to acquire data faster[13]. The board can excite up to 12 tones simultaneously over a 100 MHz bandwidth, and sampling rate of the acquisition can be varied from 500ksps up to 2Msps. The maximum number of tones is actually limited by the amount computational memory required to strongly filter the different tones in order to avoid beatings between them. The tones are generated in-phase and in-quadrature phase by two 14bits DACs, mixed with a Local Oscillator signal up to GHz frequencies, sent into the colder stage of the cryostat housing our detectors, then amplified by the cold amplifier before to be down-mixed again to electronics bandwidth and red by FPGA 14bits ADC (see Figure 2). The acquisition data process typically follows these steps:
- Make the frequency sweep to search for the resonances and calibrate the relation between frequency and phase of the excitation tone[8];
- Place the tone on the minimum of each resonance;

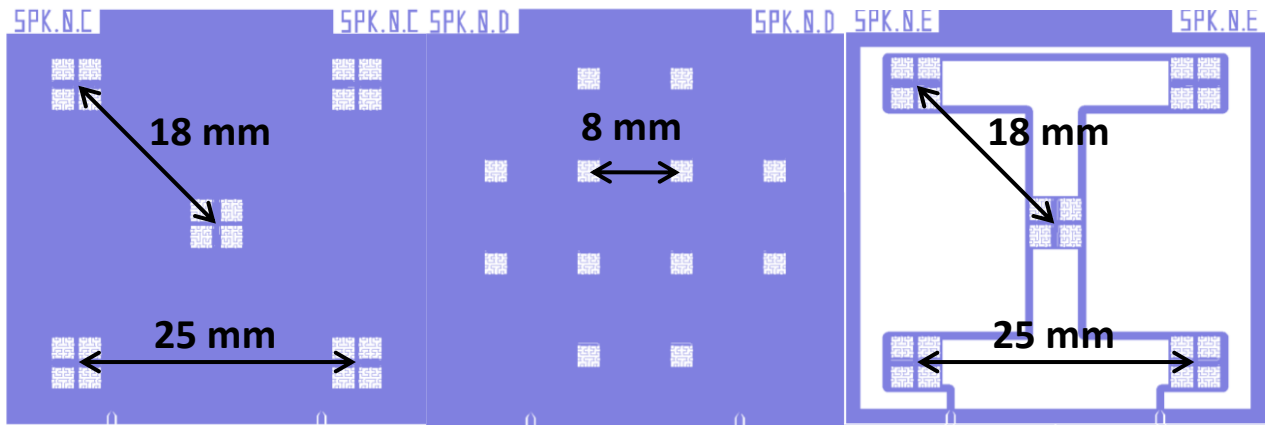

Figure 3. The dedicated SPK mask. Blue is the metallization. Each white square is where one of the Hilbert LEKID pixel is located (not visible on this scale). Starting from the left, the 'Island' array with ground plane (SPK-C), the 'Distributed' array with ground plane (SPK-D), and the 'Island' array without ground plane (SPK-E)

- Fix a threshold on the amplitude of the signal. Each tone has its specific threshold. Since the tones are placed on the resonance, when energy is absorbed in the detector the resonance will shift to lower frequencies and a positive spike in the amplitude will be observed. If the spike reaches higher than the fixed threshold, the FPGA considers it an 'event', records the data for a fixed period of time and transmits them over an ethernet connection to the acquisition PC. Each time that at least one of the tones has an event, the data are acquired for all the tones. This allows us to check for coincidences among different pixels. Furthermore, it prevents us from missing the smaller events that might be in the noise for a pixel far from the point of impact, as this same event will produce a larger effect on the nearer pixels.

### 4.3 Detector array (dedicated SPK mask)

All our detectors are designed and fabricated within our collaboration at the Institut Néel and were conceived for millimeter astronomy. Building on the NIKA experience[1,12], they are based on Lumped Element KID with an Hilbert geometry[14], fabricated with a 18 nm aluminum deposit. In order to study KID iteration with high energy particles, a dedicated mask with different focal plane geometries has been designed. First of all, in order to provide spacing in frequency domain matching the limits imposed by our fast electronics, on each array there is a smaller number of pixels with respect to a typical array designed for astronomical applications. On the other hand, to still be able to sample the whole focal plane area, the pixels are sparser on the array. The larger arrays on the mask are of the same size of the NIKA-like chips, 36x36 mm$^2$, and are the ones that have been used for the measurements.

In order to study the energy propagation in function of the distance from the point of impact of the high energy particle, two main geometries have been conceived (see Figure 3):

- Distributed, 12 pixels evenly spaced. This geometry is useful to sample an high number of distance from the impact point and then try to fit a law for the energy decay in function of the distance. We call this geometry SPK – D;
- Islands, 5 well-spaced groups of 4 very close pixels (that we call 'islands'). This allows us to monitor the effect of the events on pixels that are very near to each other and at the same time to estimated phonon velocity of propagation measuring eventual delays in the signals from different islands. This geometry has been designed either with the ground plane filling the whole surface of the array (SPK – C) or both with the most of ground plane between pixels removed (SPK – E).

This mask has been also replicated using different substrates to test the effect of the materials used on the phonons propagation. In particular, we have performed measurements on substrate with different resistivity to study the down-conversion of high energy phonons. We also tested a multilayers substrate to create a high thermal resistance to prevent these phonons from getting into the detector (see section 5).

### 4.4 High energy sources

Different high energy sources are available to study the effect of impacts on the detectors or the substrate and thus simulate the Cosmic Rays events. Our available options are $^{244}$Am and $^{241}$Cm sources, emitting alpha particles of 5.4 MeV and 5.9 MeV respectively. Another possible option is the direct use of secondary Cosmic Rays. These are essentially muons, which are produced by impacts of primary CR in the atmosphere. Each muon deposits roughly 200 keV in our substrates (see paragraph 5.4).

## 5. MEASUREMENTS

### 5.1 Use of thin membrane

The first solution that could be adopted to improve the performances of the detectors in space conditions is that of making pixel on a thin membrane. This can help in two ways: first, if a CR hits on the membrane, only a smaller fraction of its energy will be deposited, as this is directly proportional to the thickness of the material. Second, the thin membrane could prevent a part of the ballistic phonon from entering it and thus limit their propagation, although this is unlikely in the light of the results already found by Planck.

Our data once again confirm this (figure 5). The test was carried out using two pixels made on the same chip and with the same geometry. One of them was on a full Si substrate, covered with a 2 μm $Si_3N_4$ layer. Under the second one the Si had been etched away so that only the $Si_3N_4$ membrane was left. Secondary CR events were acquired for each of the 2 pixel. The presence of the $Si_3N_4$ layer made the data noisier, so that only stronger events could be observed, which reduced substantially the number of glitches observed. It is nonetheless possible to compare the rate of events in the two cases. This turns out to be essentially the same, roughly 0.1 event/min with the chosen settings, corresponding to around 4 events/s/m$^2$. The pixel on the membrane and the one on the full substrate are therefore sensitive to the same effective area, which implies that the presence of a thin membrane does not prevent the phonons from propagating into it.

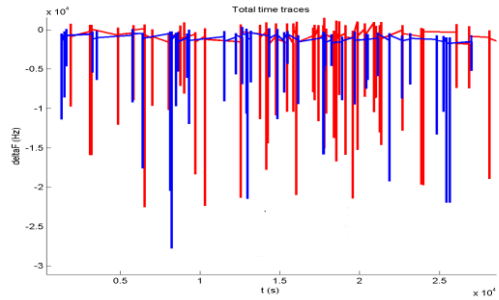

Figure 4. The total time trace of the acquisition of secondary CR events for KID on the full substrate (red) and on the membrane (blue). It can be noticed how the event rate is very similar in the two cases, of about 0.1 event/min.

### 5.2 Comparison between substrate with different resistivities

As we have mentioned before, a KID is sensitive only to the high energy phonons. A substrate with lower resistivity will have a higher number of impurities in the lattice, which can act as scattering centres for such phonons and fasten their thermalization. If it is the phonons down-conversion process that governs the time evolution of the glitches, this solution should in principle make them shorter. Furthermore, it might decrease the distance over which the ballistic phonons can propagate, thus making only a smaller part of the array affected by each impact.

We have observed the effect of secondary CR events using a SPK-E array reading 2 pixels for each 'island', for a total number of 10 pixels. On a silicon substrate of low resistivity (0.5÷1 Ohm cm), we could see that the number of coincidences was typically only 1 or 2 for each event (left histogram in figure 6). This result implies that we were able to see the same event in general only for nearby pixels and not between different islands. Yet, the noise on this array was much higher than in the case of the high resistivity ones (see figure 7), so the fact that no coincidences were observed on far away islands might be simply due to the fact that the glitches were present but hidden in the noise. This would add a non-gaussian component to the noise itself, an effect that must be definitely avoided. The low-resistivity substrate is therefore not a viable solution.

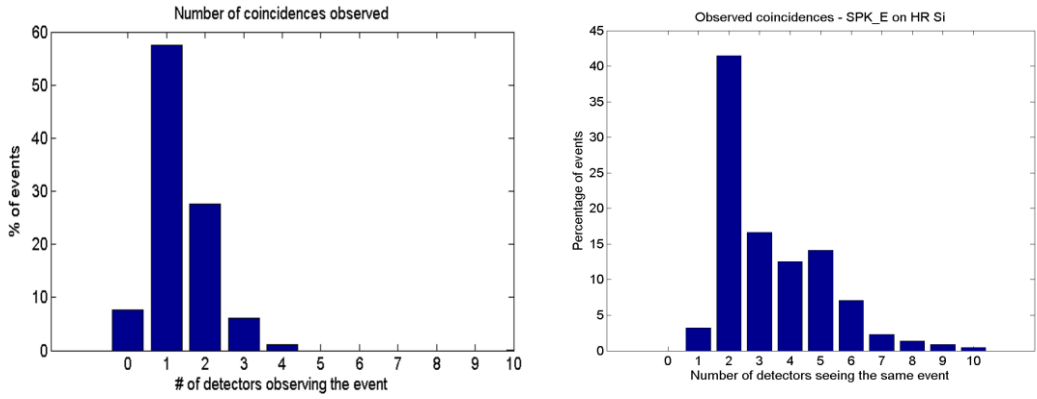

Figure 6. A comparison of the coincidences observed on SPK-E array using a low resistivity substrate (between 0.5 and 1 Ohm cm, on the left) and a high resistivity one (>1 kOhm cm, on the right). The maximum number of pixels that can see the event at the same time is the 10.

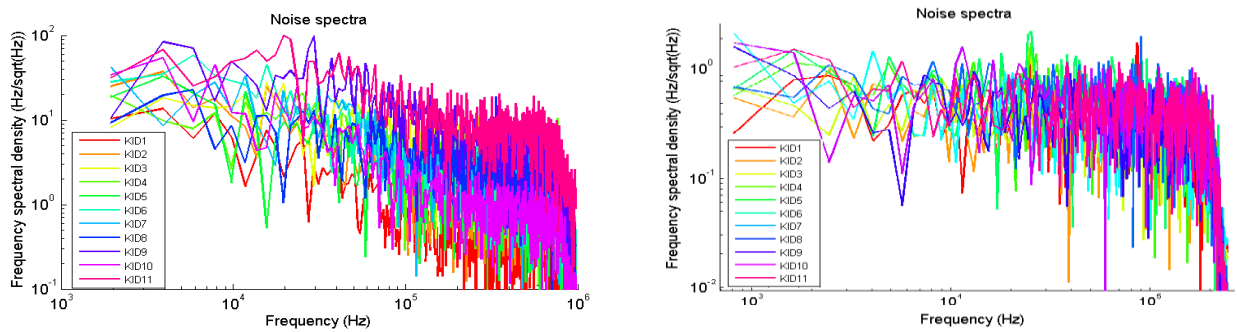

Figure 7. The noise spectra of the signal observed on SPK-E array fabricated on medium (on the left) or high resistivity substrate (on the right). The spectra were measured for 11 KID, 10 of which were used to plot the coincidences histograms. The band coverage is different as a consequence of the different sampling rate of the acquisitions. The spectra of the lower resistivity substrate show a noise level roughly 10 times higher and an evident 1/f contribution, which can easily explain the lower number of observed coincidences.

### 5.3 Measurements with secondary CR on high-resistivity substrate

In order to better understand the mechanism of propagation of the energy in the substrate, we need to minimize the noise. For this reason we performed a series of experiments on SPK arrays fabricated on high-resistivity substrates (>1 kOhm cm). First of all, we studied the effect of secondary CR on the SPK-E array. The total time trace of the array (figure 8, left plot) lets us estimate a rate of about 18 events/min (corresponding to roughly 200 events/s/$m^2$), in good agreement with the expected rate at sea level[15], assuming that we are able to sense all the particles hitting the array surface. A histogram of the number of pixel observing the same event is shown in the right plot of figure 6. It can be seen that more than 50% of the events are sensed by 3 or more pixels. Since we were reading out the signal from 2 pixels per island, this means that most impacts are observed by at least 2 islands. The minimum separation between them is of about 18mm (see figure 3), which means that each Cosmic Ray event can affect a large portion of the array. The signal is in general visible even in pixels that are more than ~18 mm away from the impact point. This confirms that some modifications are needed with respect to the currently available solutions before adopting them for space based missions.

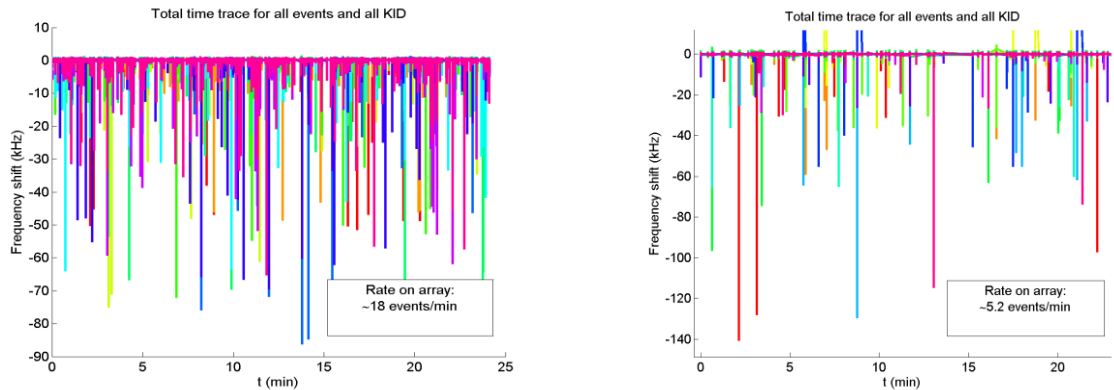

Figure 8 The total time trace of the acquisition of secondary CR events for the pixel on the SPK island geometry. The event rate observed is significantly different between the array without ground plane (SPK-E, on the left) and with ground plane (SPK-C, on the right).

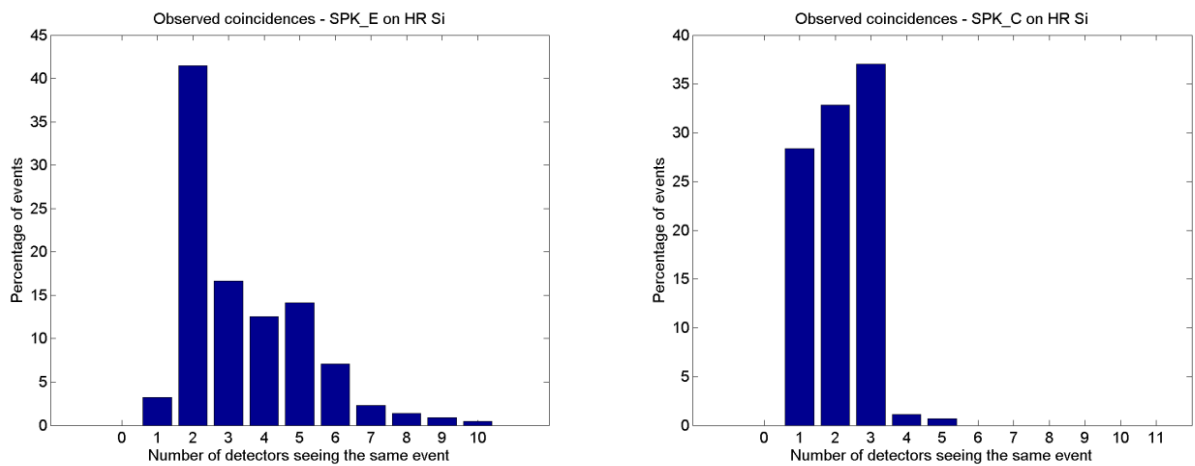

Figure 9. A comparison of the coincidences observed on SPK-E array without ground plane (on the left) and SPK-C array with ground plane (on the right). 2 KID for each island were fed for SPK-E array, while there were between 1 and 3 pixels fed per island for the SPK-C array.

One conceivable option in order to reduce the portion of array affected by CR events is that of increasing the fraction of the area that is metallized. This surface can absorb part of the phonons and enhance their energy down conversion. Therefore, we repeated the measurements on an SPK-C array, which has the same geometry of the SPK-E one, but with the ground plane filling the whole array surface. During the acquisition, between 1 and 3 tones were active for each island. The histogram in figure 9 implies that in this case we are able to see the same event only on pixels belonging to the same island. Thus, the ground plane effectively helps reducing the energy available for the detectors, likely because, as said above, it acts as an additional channel for the thermalization of the phonons energy. This is furthermore confirmed by the total time trace in the right plot of figure 8, which shows a decrease of about a factor 3.5 in the rate of observed events. This means that the effective area sensitive to secondary CR impacts is reduced by the same amount. A back-of-the-envelope calculation allows us to tell that we can detect only particles hitting the array less than 5mm away from the centre of an island. Although not yet conclusive, these results encourage us to further investigate the solution of adding metallization layers.

**5.4 Measurements with alpha sources on SPK arrays**

In order to obtain a clearer framework in which to study the energy propagation into the substrate, we need to significantly increase the statistic of our experimental data. For this reason we chose to use high-energy particle sources with a considerable radioactive activity to simulate the effect of CR events in our detectors. The energy deposited into the substrate by an alpha particle is much higher than in the case of a CR. As already said in section 2, a CR impact releases about 0.5 keV per μm traversed for a silicon target. Typically our substrates have thickness of 300μm, which leads to an average deposited energy of roughly 200 keV, depending on the angle of impact. On the other hand, an alpha particle is completely stopped on a distance of the order of few μm into a silicon substrate[16]. This means that all their energy is released into the substrate close to the impact point. Since our radioactive sources emit alpha particles of about 6 MeV, this increments the energy deposited in the substrate of a factor 30 with respect to CR hits. It can therefore be possible to detect a signal even in the furthest pixels, which in turns results in a better sampling of the function describing the phonons energy propagation. The alpha sources were mounted in front of the array, using a copper collimator with a length of about 7cm and an aperture diameter of 2.5mm. Besides providing a substantially higher events rate with respect to CR (about 1 hit per second), this solution also fixes the impact area making it possible to determine with a good approximation the distances between the impact point and the detectors. Relying on the high hit rate and deposited energy, and on the known position of the impacts, we aim to give a preliminary estimation of the damping of the signal as a function of the distance and of the time.

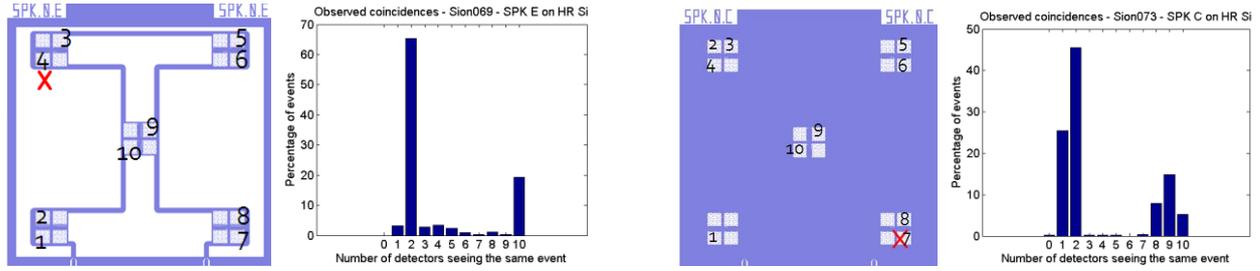

Figure 10. The geometry of the active KID with the histogram of observed coincidences on SPK arrays, with (SPK-E, on the right) and without ground plane (SPK-C, on the left) exposed to $^{244}$Am alpha source. The impact areas are indicated in the images by a red cross.

The experiments were performed with the $^{244}$Am on SPK-E and SPK-C arrays (see figure 10). The collimator was placed in front of one of the corner island, in order to maximize the distance between the hit point and the furthest island, which corresponds to about 36 mm. The total event rate was similar for the two arrays, corresponding to about 150 events/min. From the histograms in figure 10, it is clear the existence of two different populations of events. The ones giving the peak around 2 coincidences correspond to secondary X-rays generated by the interaction of the alpha particles with the copper collimator in proximity of its aperture. Since these photons have a much lower energy compared to the alpha particles, they induce a signal only in the detectors closer to the impact area. This is confirmed by the fact that these hits are always observed only by the KID of the target island. We can thus conclude that this population actually belong to another kind of events. The alpha events correspond on the contrary to the lower rate population. Their glitches can be seen by the whole array, which means that their energy is detectable at least at a distance of about 36mm, both for the array with ground plane and the one without it.

By comparing the two histograms in figure 10, we can observe a slight diminution of coincidences on the array with the ground plane. X-events are seen almost always by both the KID of the target island in the SPK-E array, while almost one third of them are seen only by KID number 7 in the SPK-C. A similar behaviour is observed for the alpha hits, which are detected by all the pixels of the SPK-E, and often by only 8 or 9 KID of the SPK-C. This is due to the fact that the glitch amplitude for some pixels of the latter array falls within their noise. This is a further evidence of the fact that the signal undergoes a stronger damping in the array with the additional metallization layer, which in our case is provided by the ground plane surface.

An example of an alpha particle event on the SPK-E array is presented in figure 11 (left plot). As highlighted by these time traces, it is clear that the KID are grouped two-by-two in families with similar characteristics. On one hand, the registered delays of the signals with respect to the detectors on the target island (KID 3 and 4) are the same for KID on the same island. A preliminary estimation gives a delay of roughly 5 μs between the target and the central island. Considering their separation of about 18mm, this results in a phonon propagation velocity of the order of 3μs/mm, in agreement with the values reported in literature for silicon[6,8]. On the other hand, pixels belonging to the same island measure similar signals. Leaving out from this analysis the glitches observed on the target KID, whose signals are highly saturated, the peak amplitudes are clearly correlated with the distance from the hit area. A plot of the signal measured by different KID normalized to the one of KID 10 is shown in figure 11 (right plot). The measured peak amplitudes are constant throughout the experiment, which demonstrates that they depend only on the distance from the impact point. Nevertheless, the amount of data collected is still not sufficient in order to provide a reliable estimation of the energy damping, principally because of the low number of sampled distances and the poor accuracy in the determination of the impact point.

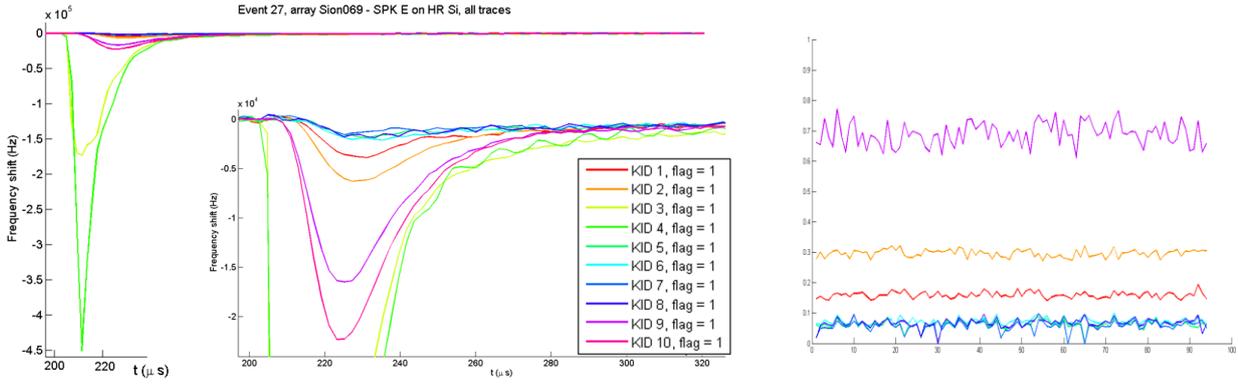

Figure 11. On the left, an example of time traces of an alpha particle event on the SPK-E array, which induces a signal in all the KID. The signals are plotted in unit of frequency shift, so that the effect is a negative glitch. In the bottom-right corner, a zoom of the time traces that evidences the presence of different families of detectors each one corresponding to a different island. On the right, the signals normalized to 10[th] KID in function of the event number. KID 3 and 4 are not considered because their signals were saturated.

The average time traces were calculated for all KID (except for those whose signal was saturated) and an exponential fit was applied. The preliminary analysis shows that an exponential with a single time constant is clearly insufficient to describe the time evolution of the signal (see figure 12). A second exponential time constant is needed in order to provide a satisfying description of the energy damping as a function of time. Although being in agreement with the literature[9], the exact nature of these two contributions needs to be confirmed by investigating the variations of these values under different quasi-particle densities environments. Moreover, the introduction of a third much longer time constant apparently enhances the quality of the fit. Nevertheless, the reliability of this additional parameter is still to be proved, as its value actually exceeds the temporal window which can be achieved by our electronics.

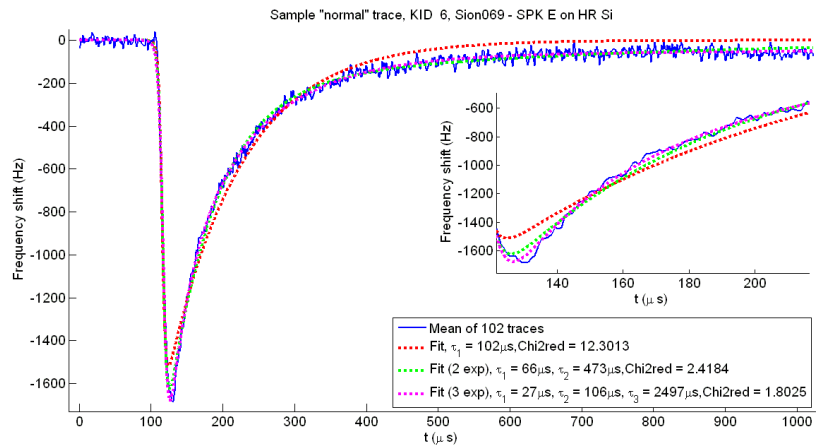

Figure 12. An example of the time trace of one of the KID of the SPK-E array exposed to the alpha source. The blue line represents the mean of 102 traces (over a total on 500 events) induced in the KID number 6 only by alpha particle hits. The exponential fits with one (red), two (green) or three (violet) time constants are plotted by the dotted lines, each one accompanied by the value of the reduced Chi-square. The variances were calculated over the pre-samples (almost 100 points).

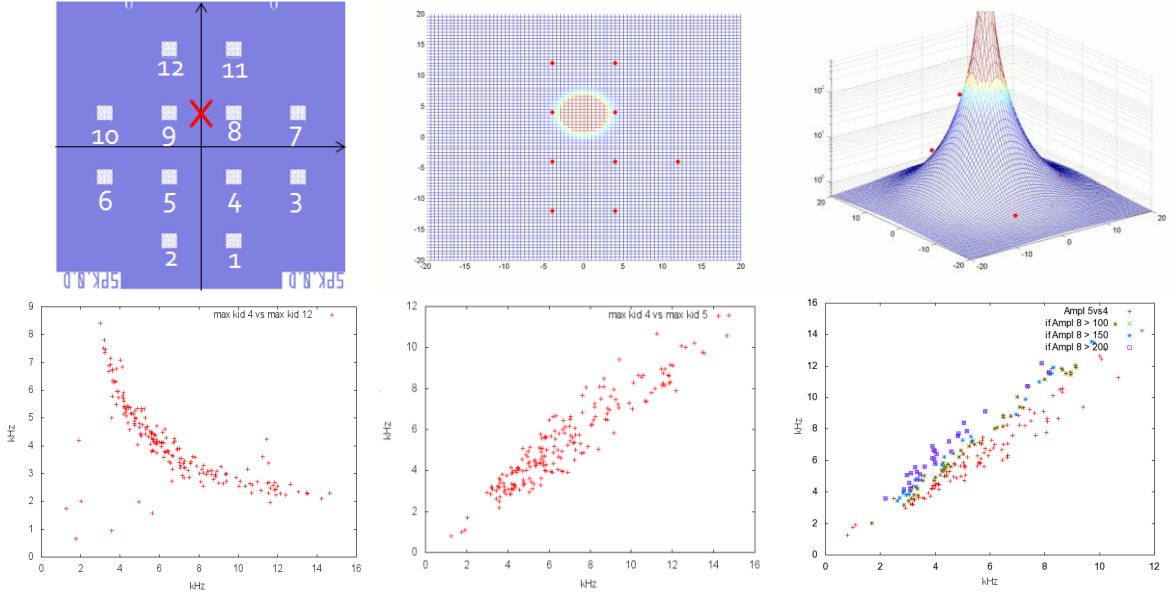

Figure 13. Measurements performed on SPK-D array exposed to a $^{241}$Cm source. In the first row: the geometry of the detectors with respect to the collimator position, highlighted by a red cross (on the right); an example of the 3-dimensional fit (on the right) and its 2-dimensional projection (in the center) applied to one alpha particle hit. The position of the KID is given onto the x-y plane, and the peak amplitude of their signals in plotted onto the z axis by the red points. KID number 6, 7 and 10 showed low quality factors and were not used in the fit. In the second row: a plot of the calibrated peak amplitude for KID 4 versus KID 12 lying on opposite sides (left plot) with respect to the hit point, and for KID 4 versus KID 5 lying on the same side of the impact point (center plot). Referring to the x-y Cartesian axes drawn in the top left plot, the variation of the y (x) coordinate of the hit position results in the linear (hyperbolic) trend, while the uncertainty on the x (y) coordinate of the impact point leads to the dispersion around this trend. This effect is highlighted by the latter graphic (right plot). Here it is repeated the KID 4 versus KID 5 signals plot, as a function of the energy peak registered in KID 8. Higher this energy, higher the ratio between the energies detected in KID 4 and KID 5, i.e. higher the slope of the linear trend.

Finally we can make a preliminary estimation of the energy propagation inside the substrate by studying the amplitude of the peak as a function of the distance $r$ from the impact point. In order to do this, we performed another experiment with $^{241}$Cm source on the SPK-D array, which was designed to better sample the surface thanks to its uniform pixel distribution. Nevertheless, the presence of the ground plane between pixels actually dampens the phonon energy, suppressing the signal which can be hidden in the noise for the furthest detectors. The acquired data are in fact not sufficient to precisely determine the function describing this attenuation. The results in the second row of figure 13 show a clear correlation of the signals coming from pixels lying on the same side of the impact point, and an anti-correlation for pixels lying on opposite sides. The linear and hyperbolic trends respectively correspond to the relationships $E_i/E_j \approx constant$ and $E_i E_j \approx constant$, where $E_i$ and $E_j$ are the energies detected by the KID i and j. These can be explained assuming that the energy damping dependency on $r$ has a power law nature. Preliminary estimates using a tridimensional fit (top right plot in figure 10) are in agreement with peak amplitudes decreasing as $r^{-4}$.

## 6. CONCLUSIONS AND FUTURE PROSPECT

We have started to investigate the suitability of KID arrays as detectors for space based missions. Our study is devoted in particular to assessing the effect of the CR hits and to limiting their impact on the final acquired data. One of the main advantages of the KID is linked to their time constants, which are typically much faster than those of bolometers. The measurements done with our detectors give time constants of the order of 100 μs, in agreement with literature. The data also hint at the existence of a slower time constant (a few ms), but its effective presence must still to be confirmed using longer data acquisitions. Assuming a dominant time constant of the order of a tenth of millisecond, a few hundreds KID array hosted on a single silicon die of ~10cm$^2$ would suffer a data loss within roughly 15%, comparable to that of one single Planck bolometer. It is worth to stress that we made the assumption that every Cosmic Ray event produces a signal in the whole array, regardless of the point of impact. Our measurements show that, over large distances (several cm), this might be in general not the case.

We have studied a series of different solutions in order to diminish the effect of CR events on the detectors far from the impact point. The first tests have included in particular the use of lower resistivity substrate and the fabrication of KID on membranes, but proved to be not viable solutions, in the first case as a consequence of the extra noise introduced, and in the latter because no variations were observed in the CR event rate.

We therefore concentrated our efforts on arrays fabricated on full wafers made of high resistivity silicon (> 1kOhm cm). The measurements in this case show a substantial difference between arrays with or without the ground plane filling the front surface of the substrate. Where present, the ground plane reduces sensibly the number of observed coincidences, leading us to conclude that its effect is to suppress more efficiently the phonon energy propagation over large distances. Our results show that a CR hit affects the whole area (3.6 cm x 3.6 cm $\approx$ 13 cm$^2$) of the SPK-C array, i.e. the one without the ground plane, whereas it only affects roughly a third of the surface of the SPK-E array. This corresponds to an effective area affected by each CR hit of about $\approx$ 1 cm$^2$ around the impact point, roughly 10 times smaller than a focal plan hosting few hundreds KID. Although this effect must be further confirmed, it already represents a possible solution for confining the effect of higher energy phonons to shorter distances.

As a last step, in order to increase the statistic and the precision of our data, we used sources of alpha particles having energy 30 times higher ($\approx$ 6 MeV) than that of the typical CR. We estimated ballistic phonon velocities in agreement with literature, and we demonstrated that their propagation into the substrate depends on the distance $r$ from the hit point. Although correctly quantifying the damping is still difficult in our experimental conditions, a first rough estimation results in suppression of the energy proportional to the 4$^{th}$ power of $r$.

A new dedicated mask will be fabricated in the next future, which uses geometries optimized to better sample the function describing the energy attenuation over distance. Furthermore, we will continue investigating the increase in the speed of the phonon energy down-conversion process which can be obtained by adding thin metal layers on the substrate. A conceivable option is that of using superconducting materials with a low energy gap on its backside. This kind of measurements and the new mask that we designed will help us provide a more quantitative description of all these effects.

## ACKNOWLEDGEMENTS

This work has been supported as part of a collaborative project, SPACEKIDS, funded via grant 313320 provided by the European Commission under Theme SPA.2012.2.2-01 of Framework Programme 7. This work is been partially funded by the CNRS under an "Instrumentation aux limites 2013" contract, by the ANR under the contract NIKA and by the CNES R&T program BSD. This work has been partially supported by the LabEx FOCUS ANR-11-LABX-0013. We are grateful to J. Baselmans and S. Doyle for the useful discussions and suggestions.